\input{epsfig.sty}
\documentstyle{l-aa}   
\newcommand{\be}{\begin{equation}}
\newcommand{\ee}{\end{equation}}
\newcommand{\bea}{\begin{eqnarray}}
\newcommand{\eea}{\end{eqnarray}}

\begin{document}
   
\thesaurus{02.01.1, 02.01.2, 02.19.1}
   
\title{On energy spectra of UHE cosmic rays accelerated in
supergalactic accretion flows}

\author{G. Siemieniec-Ozi\c{e}b{\l}o, M. Ostrowski}

\institute{Obserwatorium Astronomiczne, Uniwersytet Jagiello\'nski,
ul. Orla 171, 30-244 Krak\'ow, Poland}

\offprints{G. Siemieniec-Ozi\c{e}b{\l}o, E-mail: grazyna\@@oa.uj.edu.pl)}

\date{Received ...; accepted .. ; }   
   
\maketitle    
   
\markboth{G. Siemieniec-Ozi\c{e}b{\l}o \& M. Ostrowski}{On energy
spectra of UHE cosmic rays ...}
   
\begin{abstract}   
Some ultra-high energy (UHE) cosmic ray (CR) events may be correlated
with the Local Supercluster plane. We
consider acceleration of such particles at the large-scale accretion flows
 of matter towards such plane and/or accretion flows onto galaxy clusters.

The formed shocks and the general compression flow are expected to allow for CR
diffusive acceleration of UHE particles. For a simplified flow geometry 
we consider a stationary
acceleration of such CRs and we discuss influence of model parameters
on a particle spectral index.  We show
that the general convergent flow pattern leads naturally to very flat
proton spectra with the phase-space spectral indices $\sigma \approx
3.0$, as compared to the `canonical' shock value of $4.0$~.

\keywords{cosmic rays -- acceleration of particles -- shock waves --
accretion}
   
\end{abstract}   

\section{Introduction}

Recent studies of the ultra high energy (UHE) cosmic rays' (CRs)
generation in large supergalactic structures are supported by
presumable detection of extragalactic  magnetic fields (Kim et al.
1991) and theoretical modelling of large scale accretion flows (Ryu et
al. 1993, Bertschinger 1985). Typical energies of UHE particles $E \geq
10^{18}$eV and the remarkable change in the observed spectrum in this
energy range indicate their extragalactic origin. Additionally, the
galactic magnetic fields  in the range of a few $\mu$G, with a typical
Larmor radius of an UHE particle $>1$kpc, essentially exclude the
acceleration sites located in our Galaxy. On the other hand the sources
of particles with  $E \sim 10^{20}$ eV cannot be more distant than $50 -
100$ Mpc due to particle interactions with the microwave background
radiation.

Several authors indicate the Local Supercluster (LSC) as a main
candidate to provide the sources of UHE CRs (Stanev et al. 1995; Sigl et al. 1999).
This idea is clearly motivated by two facts. First, this is a natural site 
of all potential UHE sources. Second, there are indications (Cen et al. 1994) that the 
large-scale accretion shocks can be a generic feature of gravitational structure 
formation.

Below, after a short discussion of physical parameters typical for LSC, 
 we analyze some aspects of the
acceleration process of energetic protons at a simplified supergalactic 
and/or galaxy cluster 
structure, with the use of the diffusive acceleration model. We show
that the general convergent flow pattern leads naturally to very flat
proton spectra with the phase-space spectral indices $\sigma \approx
3.0$, as compared to the `canonical' shock value of $4.0$~. The actual
value of the index depends on a number of parameters including the
involved spatial scales of the flow, the involved velocity of accreting
flow and diffusive properties of the medium, depending
on the magnetic field structure.

\section{Physical conditions in supergalactic accretion flows}

In analytic and numerical studies of the evolution of the mass distribution 
one can see the hierarchy going from galaxy clusters up to two-dimensional 
structures called {\it walls} or {\it sheets} and {\it filaments} being at 
the intersections of such walls.
 The same picture is revealed by observations of 
the distribution of luminous matter (e.g. Lapparent et al. 1991).

Below we would like to recall the present status of the diffusion acceleration 
in large-scale shocks accompanying a structure formation in the universe. Thus, 
in the following we briefly review the constraints for the main parameters -- 
the strength of extragalactic magnetic field, and the accretion flow velocity -- 
which are required to accelerate protons to energies beyond the EeV scale.  
 Cosmic magnetic fields beyond the Galactic disk are poorly known. 
 There 
are however some observational indications for its existence in galaxy 
cluster cores as well as in their outer regions (Kronberg 1994;  
Kim et al.1991; Ensslin et al.
1998a; Valee 1990, 1997). The observation of diffuse radio emission from
 galaxy clusters provides 
evidence that magnetic fields and relativistic electrons are distributed there 
on megaparsec scales. Typical $\mu$G fields and the 100 kpc scales were 
detected by Faraday rotation measurements, but magnetic fields are still undetectable for 
larger structures. Nevertheless, recently, existence
 of large-scale magnetic field correlated with large-scale structure of
the universe is often hypothesized 
(Ryu et al. 1998; Kulsrud et al. 1997; Medina Tanco 1998).
 There 
are X-rays from the hot gas detected outside clusters
(Soltan et al. 1996), which allow 
to assume that the significant magnetic field  
 occur there on scales typical for galaxy superclusters. Unfortunately, unless 
we directly see the magnetic fields at radio or $\gamma$-ray secondaries to 
UHE CRs (which spectral shape is sensitive to cosmological magnetic fields; 
see discussion Lee et al. 1995),
they must serve 
only as a postulate.

The nonuniform extragalactic magnetic field associated with the large-scale 
filaments and sheets is supposed to be responsible for the Faraday rotation of 
extragalactic sources (Ryu et al. 1998). In contrast to the upper limit 
$B_{ext} \leq 10^{-9}$G derived from RM of quasars based on the assumption 
of magnetic field uniformity, in the case of non-uniform fields, the field strength  
 is expected to be in the range $10^{-9} - 10^{-6}$G. This high 
strength inside the cosmological walls ($\sim 0.1 \mu $G) substantially 
decreases in the surrounding voids. According to the simulations 
the field can be well ordered along the structure for several megaparsecs. The 
relatively high strength of magnetic field in the walls could be due to 
turbulent amplification associated with the  
of large-scale structure formation (Kulsrud et al. 1997).

 On the other hand, both numerical simulations
 (Ryu et al. 1993) and
theoretical modelling (Bertschinger 1985) of structure formation points
out that large scale accretion shocks must occur, when the diffuse matter
falls down to generated deep potential wells. These shocks amplify
 and order the magnetic field. The increase 
of the strength and the field coherence length is suggested to be limited
 by the energy equipartition 
state, which in the case of large scale sheets should be smaller than the value
 quoted by Ryu et al. for filaments. Thus the typical upper value of B$_{sheet}$ 
is expected to be $\sim 0.1 \mu$G at 10 Mpc coherence length scale.
 Although one cannot yet
confirm the existence of large-scale flows with direct observations it
has been suggested (Ensslin et al. 1998b) that the shocks coupled with
  galaxy clusters may be 
responsible for acceleration of electrons which we observe in the 
 so-called `cluster radio relics'. The regions of cluster relics which show diffuse radio
emission and do not coincide with any host-galaxy are treated as tracers
of accretion shock waves developing at large-scale plasma inflows onto 
galaxy clusters (Ensslin et al. 1998a). 

In spite of lack of evidence of shocks associated with sheets and filaments, 
one believes, according to hydrodynamical simulation and theoretical investigations, 
that they are formed on the border of the largest scale structure.

The properties of the considered shocks like the shock position and its velocity, $u$,  
can also be derived through numerical simulations. In particular, shocks around 
clusters are typically described by $u \sim 1000$ km/s (Kang et al. 1996). 
In the case of larger structure, one only assumes that the accretion velocity 
onto a galaxy wall should be consistent with galaxy streaming velocity. Thus, for 
 equipartition $\sim 0.1 \mu$G magnetic field the  
 accreting matter velocity  $u
\sim 400$ km/s is comparable to the characteristic turbulent velocity (Kulsrud et al.
1997). This value is consistent with the bulk flow motion of field spiral galaxies
(Giovanelli et al. 1998; Dekel 1994). Determined simultaneously by observations,
the gravitational instability theory and the numerical simulations, the coherent estimate 
of streaming motion velocities give values of $250 \pm 40 $km/s at a distance 
of 10 Mpc from LSC plane. The simulations and analytic approach suggest their increase at the 
smaller scales of order of a few Mpc (R. Juszkiewicz, private comm.). Thus the value 
of 400 km/s at the shock location seems to be reasonable. 
Below, for the discussed accretion flow/shock structures we will derive spectra 
of accelerated high energy protons with the use of a simplified one dimensional 
model.

\section{A model of stationary acceleration}

 Let us consider a 1D
steady-state symmetric model for UHE CRs acceleration (Fig.~1). 
We assume that seed particles are provided for the cosmic rays acceleration
mechanism by the galaxies concentrated
near the central plane of a flattened supergalactic structure. On
both sides this structure is accompanied by planar shock waves (Fig.~1).
For numerical estimates we locate the shocks at the distance $x_0 \sim 3$ Mpc
 from the supergalactic
plane and for the accreting matter velocity we assume 
  $u
\sim 400$ km/s.
 Then, for an order of $0.1$ $\mu$G  supergalactic 
magnetic field,
 the diffusive description can be still valid for
particles with energies reaching $10^{20}$ eV (Blasi et al. 1999; Sigl et al. 1999).
For a finite extent of an acceleration region, in order to obtain the power-law 
particles' distributions,
 one can assume 
 a particle
spatial diffusion coefficient $\kappa$ to be a constant. As one cannot
expect this simplification to hold in real objects, the obtained
solutions should be treated as approximations valid in a limited energy
range.

 The considered diffusive transport equation for cosmic ray phase space 
 distribution function, $f(x,p)$, 
can be written in the form

\be
{u(x) {\partial f \over \partial x} - \kappa {\partial^2 f \over
\partial x^2} - {1 \over 3} {\partial u(x) \over \partial x} p {\partial
f \over \partial p} = Q(x,p)} \quad ,
\ee

\noindent
where
\begin{itemize}

\item a monoenergetic source is given in the central plane $x = 0$ by
$Q(x,p)=Q_0\delta(x) \delta(p-p_0)$,

\item the velocity $u(x) = u_1(x) = - u_1 sgn(x)$ of accreting matter in both regions
upstream of the shock ($\mid x \mid > x_0$) is assumed to be constant,

\item the velocity $u(x) = u_2(x)$ in the internal region of the
structure -- downstream of both shocks ($\mid x \mid < x_0$) -- is
Linearly decreasing towards the central plane to model the plasma
compression process accompanying the structure formation inside the
supercluster. With the normalization constant $C$ defined by the shock
compression ratio $r$ -- $\mid u_2(x_0) \mid \equiv C x_0 = u_1 / r$ -- it can be
written as

$$ u_2(x)  = - C \, x \quad , $$

\item $\kappa$ is the CR diffusion coefficient. It is taken constant in
the whole space to enable an analytic solution of Eq.~1~, which is the
power-law in particle momentum.

\end{itemize}

\noindent
The free-escape boundaries are given at $x = \pm L$ :

\be
f(-L,p) = 0 = f(L,p) \quad .
\ee

\noindent
One requires continuity of both, the proton distribution $f(x,p)$ and the
differential particle flux

\be
S(x,p) \equiv -4\pi p^2 \left\{ {u_i \over 3} p {\partial f \over
\partial p} + \kappa {\partial f \over \partial x} \right\}
\ee

\begin{figure}                    
\vspace{5.5cm}
\includegraphics{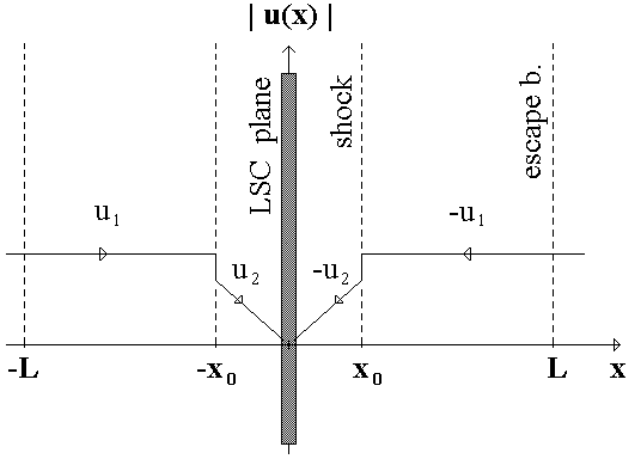}
\caption[ ]{A simplified symmetric supergalactic accretion flow model
considered in the present paper. The involved spatial positions of
accretion shocks at $\pm x_0$ and of particle escape boundaries at $\pm
L$ are indicated with dashed lines. An inflow velocity plot is superposed on
this spatial structure.} \end{figure}

\noindent
at the shocks ($x = \pm x_0$) and continuity of $f(x, p)$  in the central
plane:

\be
[f]_{x=\pm x_0} = 0 = [f]_{x=0} \quad ,
\ee

\be
[S]_{x=\pm x_0} = 0 \quad .
\ee

\noindent
At the central plane a flux discontinuity occurs due to the source
term

\be
[S]_{x=0} = -4\pi p_0^2Q_0 \delta (p-p_0) \quad .
\ee

For positive $x$ a solution of Eq.~1 can be found in a separable form
$f(x,p) = G_i(x) A(p)$, where we put $i = 1$ ($2$) for $x > x_0$ ($x < x_0$), 
through the succesive application of the above conditions. For negative
$x$ the solution is provided by the symmetry $f(-x,p) = f(x,p)$ (cf.
Fig.~1). In the upstream region, $x_0 < x < L$, we obtain

\be
G_1(x) = \exp \left [ {u_1 \over \kappa} (L-x) \right ] -1 \quad .
\ee

\noindent
In the downstream region ($0<x<x_0$) the solution is expressed by the
hypergeometric confluent function (Abramovitz \& Stegun 1965)

\be
G_2(x) = {}_1F_1 \left[ -{\sigma \over 6}, {1 \over 2}, {C x^2 \over 2\kappa}
\right] \quad .
\ee

\noindent
The resulting particle momentum spectrum for $p>p_0$ has the power-law
form

\be
A(p) \propto \left( {p \over p_0} \right)^\sigma \quad .
\ee

\noindent
In order to derive the spectral index $\sigma$, one has to match (Eq-s~4 and 5)
the above solutions at the shock. After introducing the new
dimensionless variables $\rho$ and $\psi$,

$$\rho = {Cx_0^2 \over 2\kappa} \quad , \quad \psi = {L \over x_0} \quad
, $$

\noindent
one gets a non-algebraic equation for $\sigma$,

\bea
{\sigma \over 3} \, & + & \left[ {r \over (1+r)} {\exp[2r\rho(\psi-1)]
\over \exp[2r\rho(\psi-1)] -1} \right. \nonumber \\
& - & \left. {{\sigma \over 3} \over (1+r)} {_1F_1[1-{\sigma \over 6},
{3\over2}, \rho] \over _1F_1[-{\sigma \over 6}, {1\over2}, \rho]}
\right] = 0 \quad ,
\eea

\noindent
to be solved numerically. Apart from the compression $r$ ($\equiv u_1/C
x_0$) the solution $\sigma(\rho, \psi, r)$ is parametrized by the two
additional parameters. The quantity $\rho = C x_0 {x_0 \over 2 \kappa} =
{u_2(x=x_0) \over u_{diff}(x_0)}$ may be regarded as the ratio of the
flow velocity to the CR `diffusion' velocity at the shock spatial scale. The
$\psi$ variable gives the confinement volume `size' measured in the shock
distance units.

In order to visualize the resulting spectrum inclination we have
plotted the spectral index as a function of  $\psi$ for a few values
of the compression ratio $r$ and the parameter $\rho$ (Fig.~2). For the 
 size of the particle confinement volume of order
of a few shock distances $x_0$, the spectral changes are
insignificant, provided the diffusive approach is valid $\rho \psi \sim u_1 L / \kappa \gg 1$.
 The
main observed feature is a rapid spectral index flattening to its
limiting value $\sigma = -3.0$ due to inefficient particle escape from
the `acceleration volume'. 
The shock compression is less significant in determining $\sigma$ (!).
On the other hand one
has to note its essential dependence on $\rho$, in particular for
small values of $\rho$ ($< 1$) admitting for noticeable spectral
changes.

\begin{figure}                    
\vspace{11.5cm}
\includegraphics{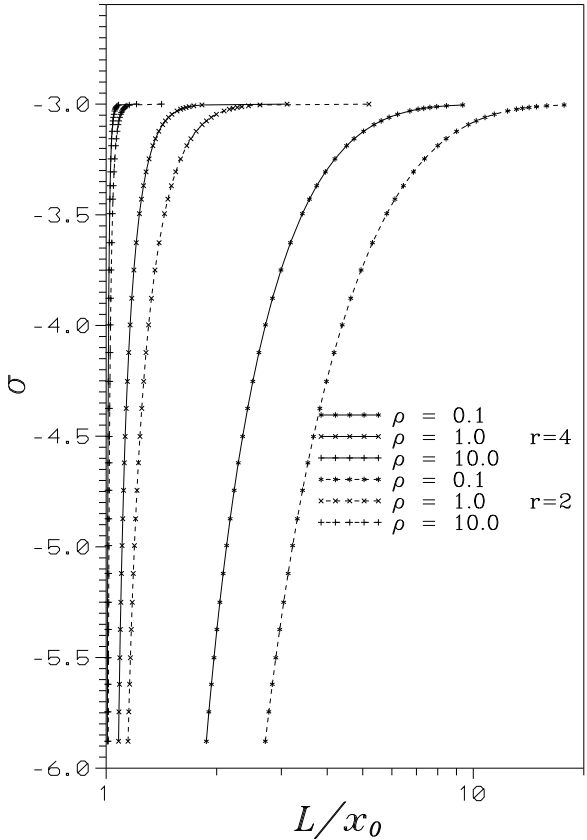}
\caption[ ]{The spectral index $\sigma$ versus the spatial scale
$L/x_0$. The results are presented for various $\rho = C x_0 {x_0 \over
2 \kappa}$ and two compressions, $r = 4$ (solid lines) and  $r = 2$
(dashed lines), as indicated in the figure.}
\end{figure}

In the asymptotic regime $\rho \gg 1$, i.e. when the particle advection
term dominates over the diffusive one at $\mid x \mid < x_0$~,  Eq.~10
reduces to

\be
\sigma \approx -3 \left[ {r \over r+1} \; {\exp \left( {u_1 \over
\kappa} (L-x_0) \right) \over \exp \left( {u_1 \over \kappa} (L-x_0)
\right) -1} + {1 \over 1+r} \right] \quad .
\ee

\noindent
In this limit, with diffusive particle
escape against the flow substantially reduced,
 the resulting index value is close to -3.0, as the
acceleration results both due to the shock compression and the plasma
convergent flow toward the supergalactic plane. On the other hand, if
$\rho \ll 1$, Eq.~10  leads to

\be
\sigma \approx
-3 \, {\exp[2r\rho(\psi-1)] \over \exp[2r\rho(\psi-1)] -1} \approx
-3 \, {\kappa \over u_1 (L-x_0)} \quad , \quad
\ee

\noindent
provided that $2 r \rho (\psi-1) \equiv u_1(L-x_0)/\kappa \ll 1$.

Requirement of the diffusive description validity yields the
lower limit for the considered $\rho$. Since the mean free path for 
proton, $l < x_0$, 
the inequality holds $\kappa < {1 \over 3} c x_0 $. Thus
$\rho > {u_1 \over c \, r} \sim {1 \over 3} 10^{-3}$, where $c$ is the
light velocity.

\section{Discussion}

The spectrum of accelerated CR protons may either become very flat or steepen 
according to the value of $\rho$ parameter. As it was seen in Fig.~2 which 
represents the solution of Eq.~10 or directly, through its asymptotic solutions, 
Eq.~11 and Eq.~12; $\rho$ controls the spectrum inclination. On the other hand 
it depends on the diffusion coefficient. The critical value of $\rho = 1$ separates 
two distinct spectra regimes. For $\rho \geq 1$ the spectral index saturates 
rapidly at the 
value of -3 (see Fig.~2) due to the compressive accretion flow predominance. 
Below the critical value, i.e. in the case of "diffusion 
velocity" greater than the inflow motions, the spectral changes can be attributed 
mainly to diffusion coefficient value. For the above considered parameters 
 of $ u_1 = 400$ km/s, $x_0
\approx 3$ Mpc, $r = 2$, the critical value of the diffusion coefficient 
corresponding to $\rho_{cr}$ is $\kappa^* = 10^{32}$ cm$^2$/s. 
For $\kappa < \kappa^*$, we get the hard spectrum with $\sigma = -3$, while for 
$\kappa > \kappa^*$ the spectrum steepens.

To estimate the maximum energy possible to achieve, let us first remind the 
rough dimensional restriction demanding the particle orbit should be smaller than 
the acceleration size $L$. In fact the realistic spectrum requires the diffusion 
length $\kappa \over u$, is smaller than $L$. For Bohm diffusion it gives the 
limitation for Larmor radius $r_g \leq 3 L (u / c) $, equivalent to $r_g \leq 50 $ kpc 
which implies $E \leq$ few $10^{18}$ eV. 
Here, we considered the
totaly chaotic magnetic field, where the Bohm diffusion gives the appropriate 
description and took for turbulent magnetic field strength $B \approx 0.1$ $\mu$G 
and $L = 10 $ Mpc.
This random magnetic field component is associated with turbulent motion which occurs 
simultaneously with streaming accretion motion at the shock vicinity, which in 
turn generates its ordered component.

The  planar 
symmetry of the model, with correlated magnetic field inside the large scale 
cosmic structure, can make the diffusion  
highly anisotropic. Therefore for $B$ aligned with supercluster plane, as 
required by Ryu at al. (1998), the cross-field 
 diffusion for a quasi-perpendicular shock should be considered. 
The minimum  diffusion coefficient in the perpendicular shocks, derived by
Jokipii (1987), is  
 $\kappa_{J} = 
3 \,{u_1 \over c}\, \kappa _B$.  
 Without 
entering into the topological characteristics of magnetic fields near the cosmic 
structure we only put here the value of critical diffusion coefficient referring 
separately to both magnetic field components and then compare the respective
 maximum energy. Above, it was clear that for entirely turbulent field, the UH 
energies can be hardly achieved.
Contrary to that, for $B$ strongly aligned 
with the structure, the diffusion can be reduced up to $({u_1 \over c})^{-1} \sim 10^3$ 
times with respect to the Bohm diffusion. Thus, for critical diffusion with $\kappa_{J} = \kappa ^*$ 
one obtains for the Larmor radius $r_g \sim 10^{25}$ cm, which corresponds to 
$E_{max} \sim 10^{21}$ eV. The latter case gives also the flatter spectrum.

The application of this model to galaxy cluster inflow is even more suitable, 
 since its spherical accretion symmetry will cause the 
acceleration process to be more efficient than in the planar case.
Adopting the same argument as used above for supercluster case, let us consider 
 the typical physical parameters
for a galaxy cluster: accretion velocity  
  $u_1 = 2 \cdot 10^3$ km/s, an upper limit for magnetic field 
  $B \approx 1$ $\mu$G and $x_0
\approx 3$ Mpc. Thus, for the Bohm diffusion one obtains for 
particle gyroradius $r_g \leq 1.8 \cdot 10^{23}$ cm. The maximum energy can reach 
the value of $10^{20}$ eV 
 and even larger for the Jokipii diffusion model.

Finally, to make sure that such scenario may serve as a viable acceleration 
process, let us confront the resulted maximum proton energy with that, when 
the energy losses are included. For UHE protons, 
 acceleration in the astrophysical shock is governed 
by the equation $dE/dt = E/\tau_{acc}$. The losses above $10^{19}$ eV 
 are mainly due to pair (e$^{\pm}$) 
and photomeson production. Both, the mean acceleration time and 
the timescale for losses, $\tau_{loss}$, has been considered in many papers. 
Their equality gives the maximum energy up to which the particles can be accelerated. 
Here, 
we use the results  
calculated and plotted by Kang et al. (1997). In their Fig.~2 the intersection point of the
 curves 
determines the maximum energy achievable in acceleration process. Taking into 
consideration the diffusion in a quasi-perpendicular case, the acceleration time 
$\tau_{acc}$ scales like (Kang et al. 1997) $\tau_{acc} \propto u^{-1} B^{-1} $,   
 to yield the maximum energy of $10^{19.6}$ eV. 

We have to note 
 that in our model the acceleration time will be smaller than the one estimated
 by  
Kang et al. (1997) and the
the maximum energy can be greater. This is due to the presence of two 
shocks associated with the compression inflow structure 
making the particle escape more difficult.  
In fact, we should consider the acceleration time estimated as 
$\tau ^{-1}_{acc} =  {\tau_{s}}^{-1} \, + \, {\tau_{c}}^{-1}$ , where 
$\tau_{c} = {p \over {{1 \over 3} p {\partial u(x) \over \partial x}}} 
={ 3 \, x_0 \over u_2} \,$ is the acceleration time scale due to adiabatic 
acceleration in the compressive flow and $\tau_s$ is the scale for the shock 
acceleration. Here, with the numerical parameters given above, we ignored 
the second term since it is comparable to the age of universe. However, it must be  
included when both $\tau_{s}$ and $\tau_{c}$ are of the same order.

\section{Conclusions}

We demonstrated that the diffusive acceleration in the accretion flow onto the
galaxy supercluster can  provide an extremely hard spectrum of
accelerated UHE protons. It is a consequence of particle confinement in the
converging flows, involving the plasma inflow towards the structure
central plane with embedded shocks. One should note that in such
convergent flows the particle acceleration process can proceed even
without shocks. Thus a possibility of a significant deviation of the UHE
CR spectral index from the often considered shock index, $\sigma_0 =
3r/(r-1)$, arises in a natural way. 
This fact should be included in modelling -- based on acceleration in
supergalactic accretion flows -- of the most energetic cosmic rays'
component observed at Earth.

Of course, the considered above a  symmetric planar model is a substantial
simplification. A divergence of the spectrum from the derived $\sigma
\approx -3.0$ may arise if particles easily escape from the structure,
$u_1 L < \kappa$, or additional particle sinks appear. The latter 
may be a result of particle escape from the accretion flow along
the supergalactic structure, extending to the sides at distances larger
or comparable to its vertical scale $L$. In both cases, the generated
spectral index -- the one in the acceleration site -- is expected
to grow with particle energy and efficiency of the acceleration process
decreases.  

\begin{acknowledgements}   
We are grateful to Dr. Z. Golda for his assistance in "Mathematica"
application and to the anonymous referee for his valuable 
comments and suggestions. This work was supported from the `Komitet Bada\'n
Naukowych' through the grants No. 2 P03D02210 (GS-O) and PB 179/P03/96/11
(MO) and its latest version through 2 P03B 112 17.
\end{acknowledgements}

\section*{References}   

\parskip=0pt   
\parindent=7mm
\noindent
Abramovitz M., Stegun I., 1965, Handbook of \par Mathematical Functions;
  Dover, New York \\
Blasi P., Olinto A., 1999, Phys. Rev.D 59, 023001 \\
Bertschinger E., 1985, ApJS, 58, 39  \\
Cen R., Ostriker J., 1994, ApJ, 429,4 \\
Dekel A., 1994, ARA\&A, 32, 371 \\
Ensslin T., Biermann P., Klein U., Kohle S., 1998a, A\&A, \par 332, 395 \\
Ensslin T., Biermann P., Klein U., Kohle S., 1998b, in \par Proc. VII Int.
  Conf. and Lindau Workshop on Plasma \par Astrophys. and Space Phys.,
  in press (astro-ph \par /9805367) \\
Giovanelli R. et al., 1998, ApJ Lett., 505, L91 \\
Jokipii J., 1987, ApJ, 313, 842 \\
Kang H., Rachen J., Biermann P., 1997, MNRAS, 286, \par 257 \\
Kang H., Ryu D., Jones T., 1996, ApJ, 456, 422 \\
Kim K., Tribble P., Kronberg P., 1991, ApJ, 355, 29 \\
Kronberg P.,1994, Rep. Prog. Phys., 57, 325 \\
Kulsrud R., Cen R., Ostriker J., Ryu D., 1997, ApJ, 480, \par 481 \\
Lapparent V., Geller M., Huchra J., 1991, ApJ, 369, 273 \\
Lee S., Olinto A., Sigl G., 1995, ApJ, 455, L21 \\
Medina Tanco G., 1998, ApJ Lett., 505, L79 \\
Ryu D., Kang H., Biermann P., 1998, A\&A, 335, 19 \\
Ryu D., Ostriker J., Kang H., Cen D., 1993, ApJ, 414, 1 \\ 
Sigl G., Lemoine M., Biermann P., 1999, Astropart. Phys., \par 10, 141 \\
Soltan A., Hasinger G., Egger R., Snowden S., \par Tr\"{u}mper J., 1996, A\&A, 305, 17 \\
Stanev T., Biermann P., Lloyd-Evans J., Rachen J., Wat- \par son A., 1995,
Phys. Lett., 75, 3056 \\
Valee J., 1990, Astron. J., 99, 459 \\
Valee J., 1997, Fund. Cosm. Phys., 19, 1 \\

\end{document}